# Giant enhancement of spin accumulation and long-distance spin precession in metallic lateral spin valves

Yasuhiro Fukuma[1*], Le Wang[1,2], Hiroshi Idzuchi[3], Saburo Takahashi[4,5], Sadamichi Maekawa[5,6] and YoshiChika Otani[1,2,3*]

[1]*Advanced Science Institute, RIKEN, 2-1 Hirosawa, Wako 351-0198, Japan*

[2]*Department of Material Physics and Chemistry, University of Science and Technology Beijing, Beijing 100083, People's Republic of China*

[3]*Institute for Solid State Physics, University of Tokyo, Kashiwa 277-858, Japan*

[4]*Institute for Materials Research, Tohoku University, Sendai 980-8577, Japan*

[5]*CREST, Japan Science and Technology, Tokyo 102-0075, Japan*

[6]*Advanced Science Research Center, Japan Atomic Energy Agency, Tokai 319-1195, Japan*

The nonlocal spin injection in lateral spin valves is highly expected to be an effective method to generate a pure spin current for potential spintronic application. However, the spin valve voltage, which decides the magnitude of the spin current flowing into an additional ferromagnetic wire, is typically of the order of 1 μV. Here we show that lateral spin valves with low resistive NiFe/MgO/Ag junctions enable the efficient spin injection with high applied current density, which leads to the spin valve voltage increased hundredfold. Hanle effect measurements demonstrate a long-distance collective $2\pi$ spin precession along a 6 μm long Ag wire. These results suggest a route to faster and manipulable spin transport for the development of pure spin current based memory, logic and sensing devices.

*e-mail: yfukuma@riken.jp; yotani@issp.u-tokyo.ac.jp

MgO based magnetic tunnel junction (MTJ) is a building block of spintronic devices such as read heads for magnetic recording and memory cells for spin random access memory[1,2]. A typical MTJ consists of two ferromagnetic electrodes separated by an insulating tunnel barrier and exhibits a large resistance change between the antiparallel and parallel alignments of the ferromagnetic electrodes, in which the spin current is accompanied by a charge current (spin-polarized current). For spin-torque switching and spin-torque nano-oscillators[3-9], a large spin-polarized current of the order of $10^{10} - 10^{12}$ A/m$^2$ needs to be applied to the junction, so that the accompanying charge current may cause adverse effects due to Joule heating and Oersted field.

Pure spin current is a flow of spin angular momentum accompanying no charge current. The nonlocal spin injection in lateral spin valves (LSVs) has proven to be an effective method to generate the pure spin current $I_S$ flowing along the slope of the spin accumulation which decays exponentially with a factor of exp ($-d/\lambda_S$) where $d$ is the distance from the interface and $\lambda_S$ is the spin-diffusion length in the nonmagnetic wire[10,11]. Subsequent spin relaxation takes place in an additional ferromagnet in Ohmic contact with the nonmagnetic wire sustaining the spin accumulation. This is so called "spin absorption" and provides an attractive means to manipulate the magnetization in magnetic nanostructures[12-15]. Therefore, it is beneficial to develop more efficient way to generate a large pure spin current as well as a large spin accumulation for advancement of novel spintronic devices utilizing lateral geometry. However, the amplitude of the voltage change $\Delta V_S$ between antiparallel and parallel alignments of the magnetization of the two ferromagnetic wires detected at the second ferromagnetic wire in LSVs is typically of the order of 1 μV. The nonlocal measurement involves no charge current flow but the spin accumulation in the vicinity of the detector. The spin accumulation is the difference in the electrochemical potential between majority and minority spins, i.e. $\Delta V_S$ is the measurable physical quantity and also decides the magnitude of pure spin current absorbed into the detector ferromagnet. Therefore efficient spin injection into the nonmagnet from the ferromagnet, being proportional to the spin valve signal $\Delta R_S = \Delta V_S / I$, and the high applied current $I$ are indispensable in realizing further enhancement of the spin accumulation[16-23].

Here, we report on the effect of interface on the spin accumulation in metallic LSVs with NiFe/MgO/Ag junctions. For achieving efficient spin injection as well as high applied current density, the NiFe/MgO/Ag junction is annealed in a nitrogen and hydrogen atmosphere. The low areal resistance of around 0.2 $\Omega\mu m^2$ of the interface MgO layer could effectively overcome the spin resistance mismatch between NiFe and Ag and leads to significant enhancement of the spin accumulation. The spin transport characteristics in the Ag nanowires have been investigated by spin valve signal and Hanle effect measurements. These experimental results are indeed consistent with the one-dimensional spin-dependent diffusion model.

Figure 1a shows a schematic illustration of our device structure. The spin polarized current is injected into an Ag wire through a MgO layer and the pure spin current flows along the slope of the spin accumulation on the right hand side in the Ag wire. The size of the NiFe/MgO/Ag junctions is 0.022 $\mu m^2$. The cross-sectional transmission electron micrograph (TEM) image is shown in Fig. 1b. The edges of the NiFe wires are covered by the MgO layer with no pinholes. Figure 1c shows the MgO thickness $t_{MgO}$ dependence of the interface resistance per 1 μm² area (resistance-area product $R_{IA}$) for the NiFe/MgO/Ag junctions at room temperature. There is no temperature dependence of $R_{IA}$. The value of $R_{IA}$ increases exponentially with $t_{MgO}$, however, the $R_{IA}$ is much lower than that for a typical tunnel junction[1,2]. The data for as-deposited and annealed NiFe/MgO/Ag junctions are respectively fitted by $R_{IA} = a\exp(bt_{MgO})$ with $a$ = 0.066 Ωμm² and $b$ = 1.41 (nm)⁻¹, and $a$ = 0.002 Ωμm² and $b$ = 0.76 (nm)⁻¹, which are shown by the solid lines in Fig. 1 c. The $R_{IA}$ is drastically decreased by annealing at 400 °C for 40 min in the 97% nitrogen + 3% hydrogen atmosphere. While it is hard to determine the composition of the MgO layer quantitatively, energy dispersive X-ray spectroscopy reveals that 6 % oxygen in MgO is decreased qualitatively after the annealing. The low $R_{IA}$ in our devices can be attributed to the oxygen vacancies[24] The nano-beam diffraction shows a typical polycrystalline pattern identified by a JCPDS card (No. 4-0829) despite the drastic change in stoichiometry of the MgO layer. Preferential growth direction is not observed in the high resolution TEM and electron diffraction measurements.

The spin valve signal of nonlocal spin injection measurements for LSV with $t_{MgO}$ = 5.5 nm and the centre-to-centre separation between the injector and the detector $L$ = 300 nm after the annealing is shown in Fig. 1d. The high and low resistances correspond to the parallel and antiparallel configurations of the two NiFe wires. The zero baseline resistance at low temperatures suggests ideal behavior for nonlocal geometry without any spurious effects due to inhomogeneous current distribution toward the detector, anisotropic magnetoresistance (AMR) and Joule heating[18,25]. Clear spin valve signals $\Delta R_S$ of 48.0 m$\Omega$ and 112.3 m$\Omega$ are respectively detected at room temperature and 10 K. In contrast, the maximum $\Delta R_S$ for as-deposited LSVs are 3.5 m$\Omega$ at room temperature and 7.5 m$\Omega$ at 10 K [26]. This fact infers that the magnitude of $\Delta R_S$ is enhanced remarkably by a factor of about 15 after the annealing.

The $R_{IA}$ and $L$ dependence of $\Delta R_S$ is shown in Fig. 2. For the NiFe/Ag junction at $L =$ 300 nm and $R_{IA} = 0.0005$ $\Omega\mu m^2$, $\Delta R_S$ at room temperature and 10 K are 2.0 m$\Omega$ and 7.5 m$\Omega$, respectively. The $\Delta R_S$ increases with increasing $R_{IA}$ up to 112 m$\Omega$ for $R_{IA} = 0.2152$ $\Omega\mu m^2$ ($t_{MgO} = 6.2$ nm), as can be seen in Fig. 2 (a). For the $L$ dependence of $\Delta R_S$ in Fig. 2 (b), $\Delta R_S$ decreases reasonably with increasing $L$ due to spin relaxation in the Ag wire. The analytical expression of $\Delta R_S$ in the presence of perpendicular field $B_\perp$ along the z-axis is obtained by solving the Bloch-type equation taking into account spin precession and (one-dimensional) spin diffusion for the LSV structure (See Supplementary Information for details),

$$\Delta R_S = \frac{4R_N^\omega \left[ \dfrac{P_I}{1-P_I^2}\left(\dfrac{R_I}{R_N^\omega}\right) + \dfrac{P_F}{1-P_F^2}\left(\dfrac{R_F}{R_N^\omega}\right) \right]^2 \left( \dfrac{\mathrm{Re}\left[\lambda_\omega e^{-L/\lambda_\omega}\right]}{\mathrm{Re}\left[\lambda_\omega\right]} \right)}{\left[ 1 + \dfrac{2}{1-P_I^2}\left(\dfrac{R_I}{R_N^\omega}\right) + \dfrac{2}{1-P_F^2}\left(\dfrac{R_F}{R_N^\omega}\right) \right]^2 - \left( \dfrac{\mathrm{Re}\left[\lambda_\omega e^{-L/\lambda_\omega}\right]}{\mathrm{Re}\left[\lambda_\omega\right]} \right)^2}, \quad (1)$$

with

$$\lambda_\omega = \frac{\lambda_N}{\sqrt{1+i\omega_L\tau_{sf}}}, \quad (2)$$

where $P_I$ and $P_F$ are the spin polarization of the currents through MgO and the ferromagnetic NiFe, respectively, $R_N^\omega = R_N \, \mathrm{Re}[\lambda_\omega/\lambda_N]$ is the spin resistance of the nonmagnetic Ag in the presence of the spin precession, $R_N$ and $R_F$, respectively, are the spin resistance of Ag and NiFe in a static condition ($B_\perp$=0), and $\lambda_N$ the spin diffusion length of Ag, $\tau_{sf} = \lambda_N^2/D_N$ the spin relaxation time, $D_N$ the diffusion constant, $\omega_L = \gamma_e B_\perp$ the Larmor frequency, $\gamma_e = g\mu_B/\hbar$ the gyromagnetic ratio, $g$ the $g$-factor and $\mu_B$ the Bohr magneton. For the analysis of the Hanle effect data, the equation (4) in ref. 28 has been widely used but is valid only for LSVs with no spin absorption into the ferromagnetic contact, e.g. tunnel or Schottky nonlocal junctions. On the other hand, the equation (1) in this study takes into account the contribution of this spin absorption to the $L$ dependence and the Hanle effects, thus enabling us to analyze $\Delta R_S$ in a unified manner[27]. Therefore, the experimental data in Fig. 2 are fitted to the equation (1) by adjusting parameters, $P_I$, $P_F$, and $\lambda_N$ with setting the value of the spin diffusion length of NiFe, $\lambda_F = 5$ nm, reported by Dubois *et al.*[29] Then we obtain $P_I = 0.42$ and 0.44, $P_F = 0.3$ and 0.35, $\lambda_N = 300$ nm and 1100 nm at room temperature and 10 K, respectively. The spin diffusion length and the interface spin polarization corresponding to the spin injection efficiency of the junction are substantially improved by the annealing; for as-deposited LSVs, $P_I$ and $\lambda_N$ are respectively 0.11 and 270 nm at room temperature, and 0.11 and 550 nm at 10 K[26]. Moreover $\Delta R_S$ for the as-deposited LSVs is drastically decreased when $t_{MgO} > 1$ nm due to the spin relaxation in the MgO layer[26], whereas in the present study the $\Delta R_S$ increases monotonically with $t_{MgO}$. This implies that spin transport characteristics of the NiFe/MgO/Ag junction are much improved. The experimental results are indeed consistent with the one-dimensional spin transport model. For the NiFe/Ag junction, injecting spin current relaxes at the interface because of the spin resistance mismatch between $R_{Ag} = \rho_N \lambda_N / t_N w_N = 0.89$ $\Omega$ and $R_{NiFe} = \rho_F \lambda_F / w_F w_N = 0.08$ $\Omega$ for $T = 10$ K, where $t_N$ and $w_N$ are the thickness and the width of the nonmagnetic wire, respectively, and $w_F$ is the width of the ferromagnetic wire. When $R_I$ reaches to the value of $R_{NiFe}$, $\Delta R_S$ starts to increase because a backflow of the injected spin current into NiFe, which may also be spin absorption, starts to be suppressed. Then, $\Delta R_S$ increases with $R_I$ and takes a maximum when $R_{IA} \sim 0.2$ $\Omega\mu m^2$ ($R_I \sim 9$ $\Omega$). This means that the interface resistance should be ten times larger than $R_N$ to overcome the spin resistance mismatch between the ferromagnetic and nonmagnetic metals.

To gain more insight into the spin transport characteristics in the Ag nanowire, the Hanle effect measurements[30,31] were performed. Figure 3 shows the modulated $\Delta R_S$ due to spin precession as a function of the perpendicular magnetic field for LSVs with NiFe/MgO/Ag junctions. For the large perpendicular field, the magnetization of NiFe tends to align toward the field direction, causing undesired change in $\Delta R_S$, the value of which is therefore corrected by using the information of magnetization process measured by AMR. The Hanle effect measurements are started from parallel or antiparallel configuration of the magnetization of the two NiFe wires. For $L = 2$ $\mu m$, the two curves cross each other at around 0.27 T of which spin precession angle is $\pi/2$. In contrast, the Hanle effect measurements for Al based LSVs with the same injector-detector separation exhibit a clear $3\pi/2$ spin precession[32], inferring much faster spin diffusion in Ag than Al. The value of $D_{Ag}$ estimated from the Einstein relation $\sigma_N = e^2 N_N D_N$, where $e$ is the electron charge, $N_{Ag} = 1.55 \times 10^{22}$ (eVcm$^3$)$^{-1}$ is density of state at the Fermi energy[33] and $\sigma_{Ag} = 8.20 \times 10^7$ ($\Omega$m)$^{-1}$ is the conductivity, is $3.30 \times 10^{-2}$ m$^2$/s, which is indeed about ten times larger than that of Al ($D_{Al} = 4.3 \times 10^{-3}$ m$^2$/s)[32]. Such fast diffusive spin transport may be suitable for faster memory, logic and sensing applications. However this fast spin diffusion hinders a large precession in the LSV with the small injector-detector separation because of short traveling time. Therefore, we fabricated NiFe/MgO/Ag LSV with $L = 6$ $\mu$m, which is much longer than the spin diffusion length. The clear spin precession up to almost $2\pi$ rotation with large amplitude is observed, as shown in Fig. 3. The experimental data are also fitted to equation (1), and the best-fit parameters $P_I$ and $\lambda_N$ are very close to the results obtained from the $R_{IA}$ and $L$ dependent measurements shown in Fig. 2. The Hanle effect measurements are known to determine the transverse spin relaxation time $T_2$, whereas the separation dependence measurements do the longitudinal spin relaxation time $T_1$. In a non-magnetic metal, $T_1 = T_2$ is predicted from analytical calculation[31,34] and is confirmed experimentally only for LSVs with $L \sim \lambda_N$[32]. Our experimental results clearly show that $T_1$ and $T_2$ are almost identical even for $L > \lambda_N$.

Finally we stress here that the low resistive NiFe/MgO/Ag junction offers a big advantage for realizing large spin accumulation in the Ag nanowire. As can be seen in the $R_{IA}$ dependence of $\Delta R_S$ in Fig. 2a, theoretical values of $\Delta R_S$ at room temperature and 10 K reach the values of saturation, 55 m$\Omega$ and 131 m$\Omega$, respectively. Note that even though the interface resistance is as low as 0.2 $\Omega\mu$m$^2$, $\Delta R_S$ shows 51 m$\Omega$ at RT and 112

m$\Omega$  at 10 K, which are close to the saturation values. Such a low $R_I$ junction allows increasing the applied current density with keeping high spin injection efficiency, as shown in Fig. 4. The large spin valve voltage over 220 μV is only realized by applying high bias current of 3.5 mA at the injector, although $\Delta R_S$ decreases due to Joule heating. The decrease of $\Delta R_S$ in high bias currents may be attributed to the spin relaxation via phonon in the Ag nanowire because the value of $P_I$, which is proportional to the spin injection efficiency into Ag through MgO, is insensitive to the temperature, as discussed for the fitting data in Fig. 2. We should note here that this spin valve voltage is much larger than that so far reported for LSVs not only for metals[16-26] but also semiconductors such as Si and GaAs[35-38] because of the efficient spin injection for the NiFe/MgO/Ag junction. There is still a possibility to further increase the spin accumulation in LSVs as CoFe/MgO[1] and CoFeB/MgO[39] interfaces in MTJs show the spin injection efficiency better than for NiFe/MgO. In this way, the low resistive MgO based junction will not only lead to efficient control of the magnetization and the domain wall by using pure spin current[12-15] but also accelerate progress towards developing novel multi-terminal spintronic devices in lateral geometry such as spin-logic devices[40].

**METHODS**

Lateral spin valves with clean NiFe/MgO/Ag interfaces are prepared on a Si/SiO$_2$ substrate without breaking vacuum by means of shadow evaporation using a suspended resist mask. The resist mask pattern consisting of 500 nm-thick methyl-methacrylate (MMA) and 50 nm-thick polymethyl methacrylate (PMMA) is fabricated by e-beam lithography. All the layers are e-beam deposited in an ultra-high vacuum condition of about 10$^{-7}$ Pa, but the ferromagnetic and nonmagnetic materials are deposited separately in the interconnected two different evaporation chambers to prevent magnetic impurities into the nonmagnetic nanowire, which degrade the spin diffusion length. Firstly, the injector and detector NiFe wires 140 nm in width and 20 nm in thickness are obliquely deposited at a tilting angle of 45° from substrate normal. Secondly, the interface MgO layer with various thicknesses ranging from 0 to 6 nm is deposited on a 20×20 mm substrate at the same tilting angle of 45°. Thirdly, the Ag wire 160 nm in width and 50 nm in thickness is deposited normal to the substrate. Finally, 3 nm thick capping MgO layer is deposited to prevent surface contamination of the devices. After the liftoff process, the LSV devices are annealed at 400 °C for 40 min in an N$_2$ (97%) + H$_2$ (3%) atmosphere.

The interface resistance is determined by a typical four-terminal method at the cross junction. The nonlocal measurements with low bias current are carried out by conventional current-bias lock-in technique. The ac current with the amplitude of 200 μA and the frequency of 79 Hz is used for the lock-in detection. The magnetic field is applied parallel to the NiFe wires for the spin valve measurements. The switching field of two NiFe wires is controlled by attaching a large domain wall reservoir at the edge. To measure the bias current dependence of the nonlocal signal, the positive dc current with the duration of 200 μs is applied at the injector (Ag (+) and NiFe (-)) and then the corresponding dc voltage is measured at the detector by using a nano-voltmeter (Ag (+) and NiFe (-)). There is no change of the magnitude of $\Delta R_S$ depending of the sign of the current. For the Hanle effect measurements, the perpendicular field along the z-axis in Fig. 1a is applied. The applied field direction is carefully controlled to rule out the misalignment which causes the in-plane field component which switches the magnetization of the NiFe wires during the measurements in a high magnetic field.

**Acknowledgements**

This work is partly supported by Grant-in-Aid for Scientific Research in Priority Area 'Creation and control of spin current' (No. 19048013) from the Ministry of Education, Culture, Sports, Science and Technology, Japan.

**Author contributions**

Y.F., L.W., and H. I., designed the experiments, fabricated devices and performed analysis. S. T. and S. M. developed the theoretical analysis. Y. O. planned and supervised the project. All authors discussed the results and commented on the manuscript.

Figure Legends

Figure 1. **Sample structure and representative nonlocal spin valve signal. a**, A schematic diagram of lateral spin valves. The NiFe wires are 140 nm in width and 20 nm in thickness. The Ag wire is 160 nm in width and 50 nm in thickness. The centre-to-centre separation $L$ between the injector and the detector changes from 0.3 μm to 6 μm. Spin polarized current $I$ is injected along the arrow on the left hand side in the Ag nanowire. Pure spin current $I_S$ diffuses in the other side of the Ag nanowire and then the spin accumulation is detected at the detector. **b**, Cross-sectional TEM image of the NiFe/MgO/Ag junction. **c**, MgO thickness dependence of interface resistance per 1 μm$^2$ area at room temperature. The as-deposited values are obtained from our previous data reported in ref. 26. **d**, Nonlocal spin signal as a function of magnetic field for LSV with NiFe/MgO(5.5 nm)/Ag junctions and $L$ = 0.3 μm. The magnetic field is applied along the NiFe wires.

Figure 2. **Interface resistance and injector-detector separation dependence of nonlocal spin valve signal. a**, Spin valve signal as a function of interface resistance of the NiFe/MgO/Ag junction at room temperature and T = 10 K. The injector-detector separation $L$ is fixed at 0.3 μm. The solid lines are the fitting curves using equation (1). **b**, Spin valve signal as a function of injector-detector separation at room temperature and T = 10 K. The thickness of the MgO layer for $R_{IA}$ = 0.0005 Ωμm$^2$, 0.0137 Ωμm$^2$ and 0.1245 $\Omega\mu m^2$ is 0, 2.6 nm and 5.5 nm, respectively. The solid lines are the fitting curves using equation (1) with the same parameters used in Fig. 2 a.

Figure 3. **Spin precession measurements by using Hanle effect.** Spin valve signal as a function of perpendicular field for LSVs with NiFe/MgO/Ag junctions with different injector-detector separation $L$ at T = 10 K. The arrows indicate the relative magnetization configuration of two NiFe wires. The solid lines are the fitting curves using equation (1) with diffusion constant $D_{Ag}$, spin diffusion length $\lambda_{Ag}$ and spin polarization $P_I$ listed in the inset.

Figure 4. **Spin valve signal and voltage for various NiFe/MgO/Ag junctions.** Spin valve signal and voltage as a function of applied bias current at the injector at T = 10 K. The injector-detector separation is 0.3 μm and the thickness of the MgO layer for $R_{IA}$ = 0.0005$\Omega\mu m^2$, 0.0032 $\Omega\mu m^2$, 0.0137 $\Omega\mu m^2$ and 0.2152 $\Omega\mu m^2$ is 0, 1.2 nm, 2.5 nm and 6.2 nm, respectively.

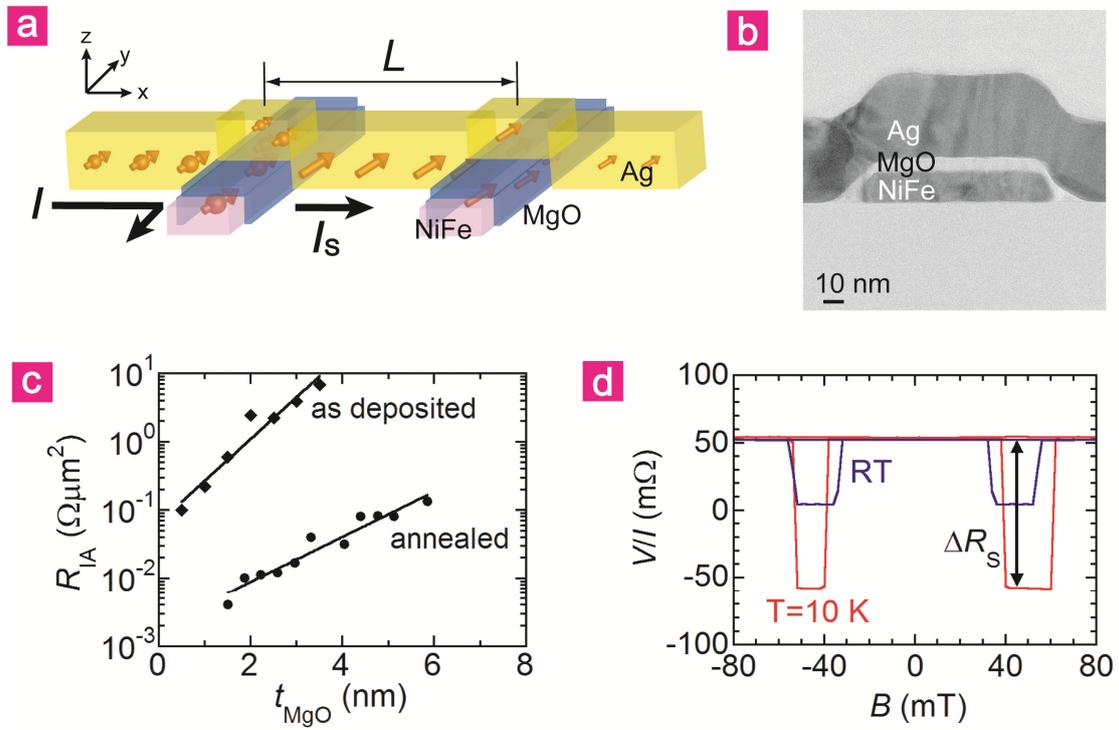

Figure 1, Fukuma *et al*

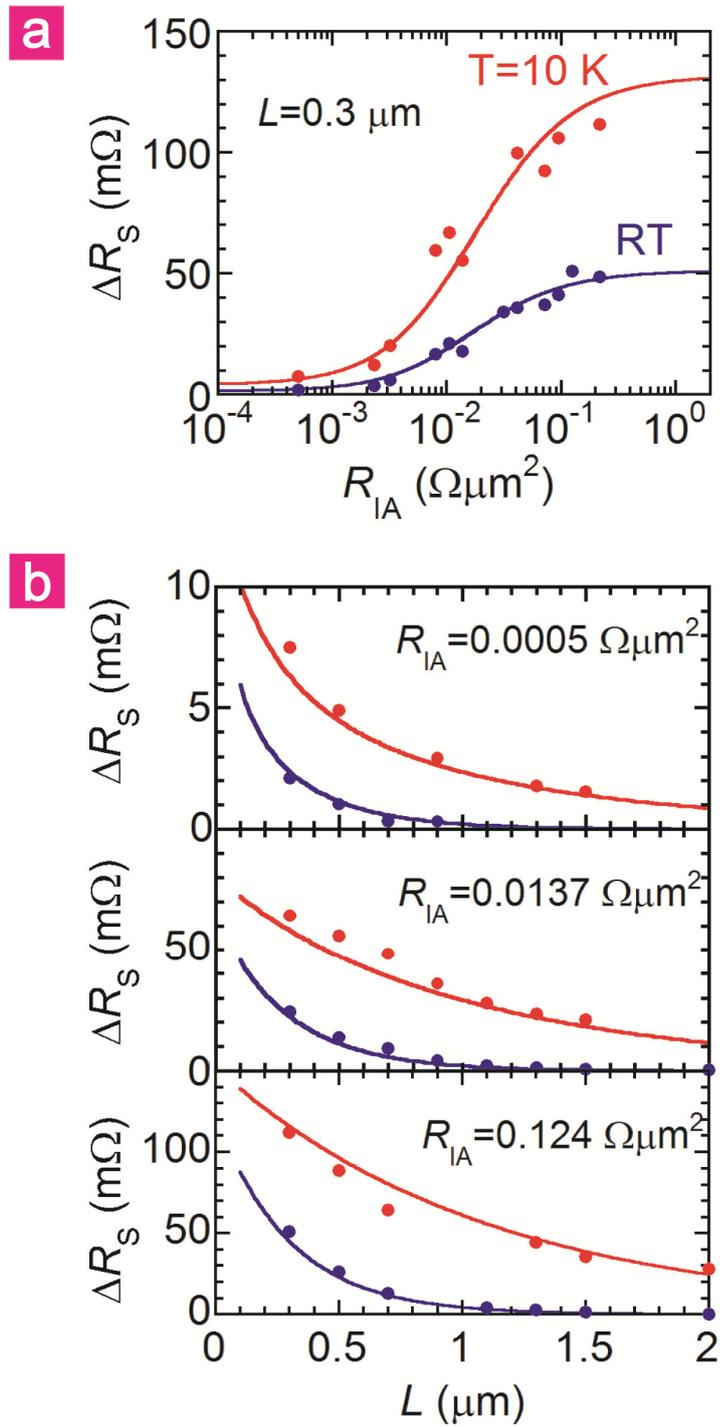

Figure 2, Fukuma *et al*

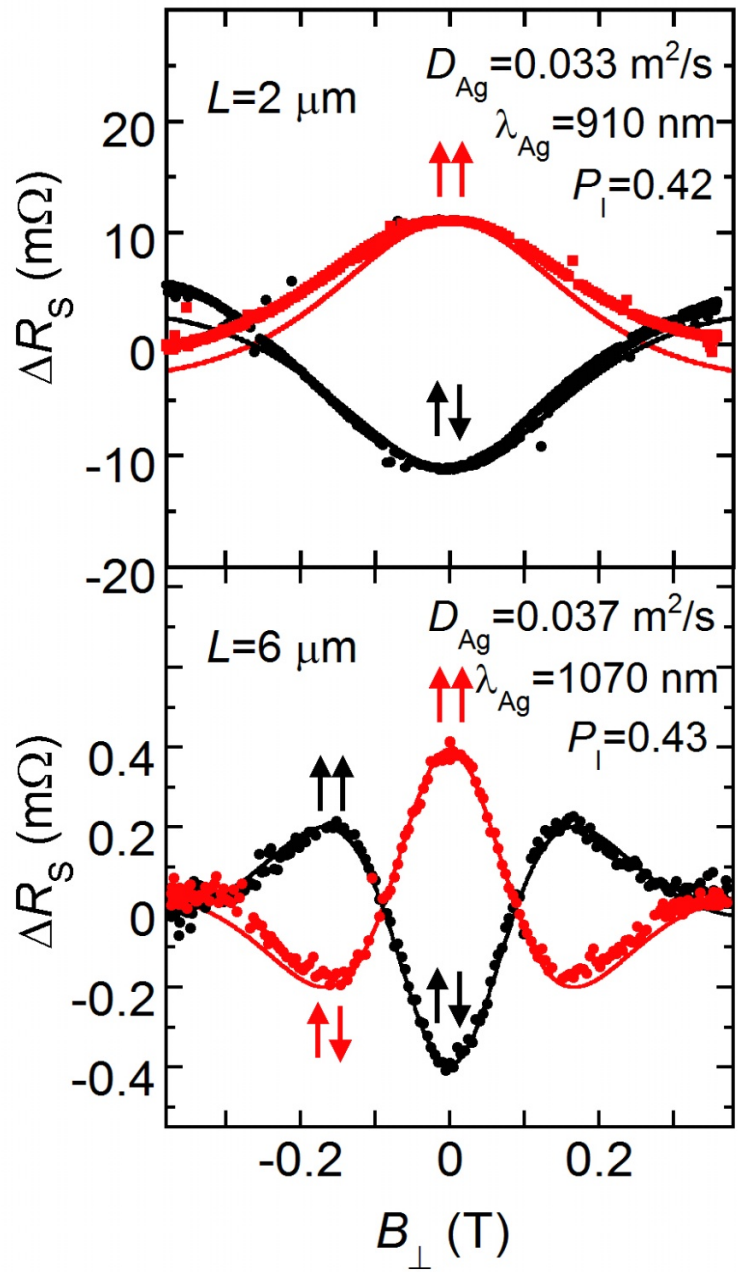

Figure 3, Fukuma *et al*

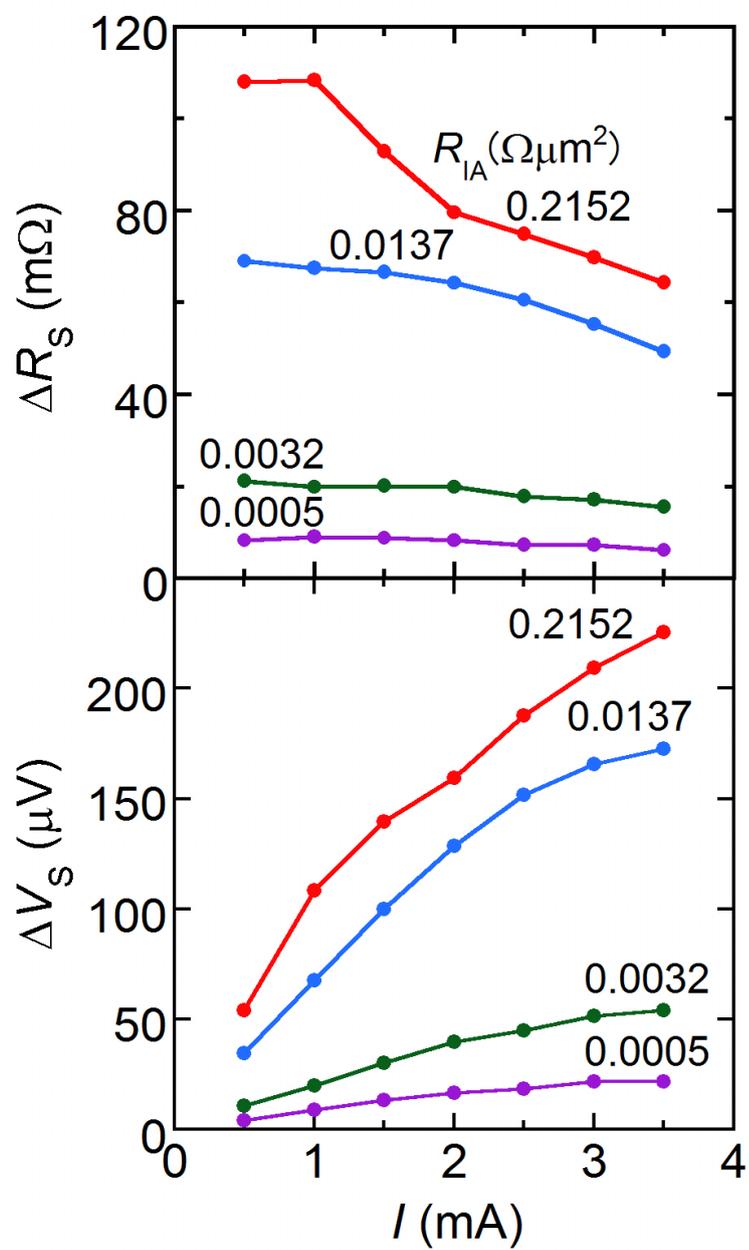

Figure 4, Fukuma *et al*